\documentclass[a4paper,fleqn]{cas-sc}

\usepackage[authoryear,longnamesfirst]{natbib}

\usepackage{xcolor}
\usepackage{caption}
\usepackage{multirow}
\usepackage{tabularx}
\usepackage{ragged2e} % for better text alignment
\usepackage{booktabs} % for nicer horizontal lines
\usepackage{enumitem}
\usepackage{mdframed}
\usepackage{graphicx}
\usepackage{csquotes}
\usepackage{adjustbox}
\usepackage{svg}
\usepackage{tikz}
\usepackage{siunitx}
\usepackage{pdflscape}
\usepackage{subcaption}
\usepackage{longtable}
\usetikzlibrary{shapes, arrows, positioning}

\usepackage{colortbl}
\definecolor{lightgray}{gray}{0.9}
\definecolor{lightgrey}{RGB}{230,230,230}
\definecolor{lightpink}{RGB}{255,245,245}
\definecolor{lightblue}{RGB}{204,224,255}
\usepackage{amsmath}
\def\tsc#1{\csdef{#1}{\textsc{\lowercase{#1}}\xspace}}
\tsc{WGM}
\tsc{QE}
\tsc{EP}
\tsc{PMS}
\tsc{BEC}
\tsc{DE}
\begin{document}
\let\WriteBookmarks\relax
\def\floatpagepagefraction{1}
\def\textpagefraction{.001}

% % Short title
 \shorttitle{Multi-Stuttered Speech Classification via Encoder based Whisper}

% % Short author
\shortauthors{H. Ameer et~al.}

\title [mode = title]{Optimizing Multi-Stuttered Speech Classification: Leveraging Whisper's Encoder for Efficient Parameter Reduction in Automated Assessment}                      
% \tnotemark[1,2]

% \tnotetext[1]{This document is the results of the research
%    project funded by the National Science Foundation.}

% \tnotetext[2]{The second title footnote which is a longer text matter
%    to fill through the whole text width and overflow into
%    another line in the footnotes area of the first page.}

%%\author[1,3]{CV Radhakrishnan}[type=editor,
%                        auid=000,bioid=1,
 %                       prefix=Sir,
%                        role=Researcher,
%                        orcid=0000-0001-7511-2910]

\author[inst1]{Huma Ameer}

\affiliation[inst1]{organization={School of Electrical Engineering and Computer Science (SEECS)},%Department and Organization
            addressline={National University of Sciences and Technology (NUST)}, 
            city={Islamabad},
            postcode={44000}, 
            country={Pakistan}}
            
\credit{Conceptualization of this study, Methodology, Software}

\fnmark[1]
\author[inst1]{Seemab Latif}
\cormark[1]
\ead{seemab.latif@seecs.edu.pk}
\credit{Conceptualization of this study, Methodology, Software}
\author[inst1]{Mehwish Fatima}
\credit{Conceptualization of this study, Methodology, Software}
\cortext[cor1]{Corresponding author}

%\ead[url]{www.cvr.cc, cvr@sayahna.org}

\credit{Conceptualization of this study, Methodology, Software}

\begin{abstract}
The automated classification of stuttered speech has significant implications for timely assessments providing assistance to speech language pathologists. Despite notable advancements in the field, the cases in which multiple disfluencies occur in speech require attention. We have taken a progressive approach to fill this gap by classifying multi-stuttered speech more efficiently. The problem has been addressed by firstly curating a dataset of multi-stuttered disfluencies from open source dataset SEP-28k audio clips. Secondly, employing Whisper, a state-of-the-art speech recognition model has been leveraged by using its encoder and taking the problem as multi label classification. Thirdly, using a 6 encoder layer Whisper and experimenting with various layer freezing strategies, a computationally efficient configuration of the model was identified. The proposed configuration achieved micro, macro, and weighted F1-scores of 0.88, 0.85, and 0.87, correspondingly on an external test dataset i.e. Fluency-Bank. In addition, through layer freezing strategies, we were able to achieve the aforementioned results by fine-tuning a single encoder layer, consequently, reducing the model’s trainable parameters from 20.27 million to 3.29 million. This research study unveils the contribution of the last encoder layer in the identification of disfluencies in stuttered speech. Consequently, it has led to a computationally efficient approach, 83.7\% less parameters to train, making the proposed approach more adaptable for various dialects and languages. 

\end{abstract}

% \includegraphics{figs/grabs.pdf}

\begin{highlights}
\item Curated a multi-stuttered disfluencies using SEP-28k and FluencyBank datasets, enhancing them by manually labeling speakers and strategically concatenating audio to simulate multi-stuttered speech.
\item Employed the Whisper model's encoder and reframed the problem as a multi-label classification for multi-stuttered speech. Achieved high micro, macro, and weighted F1-scores of 0.88, 0.85, and 0.87, respectively on an external test dataset, demonstrating significant advancements in handling multiple disfluencies.
\item Achieved a substantial reduction in trainable parameters from 20.27 million to 3.29 million by employing strategic layer freezing, resulting in a computationally efficient model adaptable to various dialects and languages.
\end{highlights}

% Keywords

\maketitle

\section{Introduction}
\label{sec:introduction}
\begin{figure}[h]
    \centering
    % Option 1: Using the `svg` package
  %  \includesvg[width=\linewidth]{whisper.svg}
    \includegraphics[width=\linewidth]{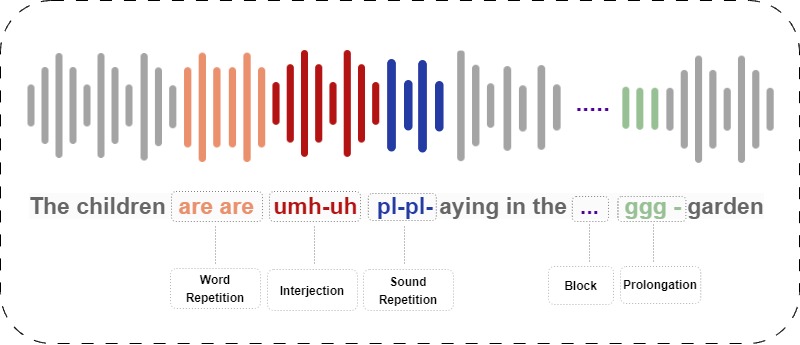}
  
    \caption{An Illustrative Overview of Different Types of Disfluency in Stuttered Speech}
    \label{fig:stutter}
\end{figure}

Stuttering, a fluency disorder is a disruption in the normal flow of speech \cite{shipley2019assessment}. People with this disorder exhibit various types of disfluencies in their speech \footnote{\url{https://www.asha.org/public/speech/disorders/stuttering/}}.
The most common disfluencies include blocks, interjections, sound repetition, word repetition, and prolongation  \cite{shipley2019assessment} which are demonstrated in Figure \ref{fig:stutter}. Stuttering affects approximately 1\% of the global population \cite{yairi2013epidemiology}.

In this era of Artificial Intelligence, automated systems are assisting professionals in various domains i.e. medical, finance, education etc. Nevertheless, in the field of psychology, particularly for Speech Language Pathologists (SLP), the assessment of stuttering patients still requires considerable cognitive effort and is time-intensive. Conventionally, SLPs have to manually identify the disfluencies in speech, count them and calculate parameters like speech rate, stuttered words per minute etc \cite{shipley2019assessment}. Consequently, the outcome is dependent on the expertise of SLPs. This creates a need for an automated tool that can assist SLPs in evaluating stuttering patients.
Although there have been numerous studies in this area, it still remains an evolving field. Since audio data is the most accessible source, most of the studies leverage it for research in the automated classification of disfluencies in stuttered speech \cite{psu3, sheikh2021stutternet}. Lately, deep learning, specifically transformer-based models has emerged as a suitable choice for this task due to the complex nature of stuttered speech \cite{bayerl2023classification, sebastian2022detecting}. Whisper \cite{radford2023robust}, a weakly supervised transformer-based model developed by Open AI \footnote{Open AI Whisper: \url{https://openai.com/research/whisper}} has proven to be a more resource-efficient and generalized model for automated classification of stuttered speech \cite{ameer2023whisper}. However, the study presented by \cite{ameer2023whisper} only deals with single stuttering, whilst in most real world scenarios, more than one disfluencies can co-occur. Although, studies presented by \cite{jouaiti2022multi,bayerl2023stutter} have addressed this issue by acquiring Wav2vec2.0 \cite{baevski2020wav2vec} and framing it as a multi-label classification for handling multi-stuttered speech instances, still the domain needs more work.

\subsection{Research Gaps and Objectives}

Despite significant advancements in disfluency classification, several gaps remain in the classification of multi-stuttered disfluencies. The aforementioned studies lack in terms of speaker-exclusive dataset splits, high annotator agreements, enough multi-stuttered data, and data quality. This study aims to address these gaps, and additionally, it also focuses on reducing the trainable parameters and finding the optimized configurations of the model. The goal is to have a resource-efficient solution and a generalized model for multi-stuttered use cases.

\subsubsection{Curating a Comprehensive Multi-Stuttered Dataset:}

\textbf{Objective:} To curate a quality dataset of multi-stuttered speech by combining and enhancing existing open-source datasets.

\textbf{Contribution:} We have curated a multi-stuttered dataset using an open-source dataset known as Stuttering Events in Podcasts (SEP-28k) and FluencyBank. The speakers were labeled manually, and then the audio of the same speakers was strategically concatenated to create multi-stuttered samples.

\subsubsection{Automated Classification of Multi-Stuttered Disfluencies:}

\textbf{Objective:} To implement a strategy for the classification of multi-stuttered disfluencies.

\textbf{Contribution:} The primary objective of this study is the automated classification of multi-stuttered disfluencies. This aspect contributes to the advancement of the field in real-world scenarios, and for that, we have leveraged the encoder-only Whisper model.

\subsubsection{Model Optimization for Real-World Applications:}

\textbf{Objective:} To optimize the model without sacrificing classification performance.

\textbf{Contribution:} By limiting the learnable parameters of the model without compromising its performance, we optimized model efficiency, making it more suitable for real-world applications. Specifically, we froze parameters including the feature extractor and encoder layers of Whisper.

This paper is divided into 5 sections: Section~\ref{sec:introduction} briefly discusses the essential background and aims of our research. Section ~\ref{sec:related work} presents a literature review highlighting the research studies related to multi-label classification for stuttered speech. Next, in the methodology section ~\ref{sec:methodology},  we have comprehensively discussed the architecture of the model followed by experimental design and configuration. The results and discussion section ~\ref{sec:results and discussion}, showcases the findings and insights of the results. Finally, section ~\ref{sec:conclusion} summarizes the key findings and future research directions.
\section{Related Work}
\label{sec:related work}
Disfluencies in stuttered speech are intricate in nature and have been addressed through various techniques applied to audio data. The acquisition of transformer-based models has revolutionized multiple domains, including computer vision \cite{parvaiz2023vision}, natural language processing \cite{patwardhan2023transformers}, and speech processing \cite{latif2023transformers,nowakowski2023adapting}. In terms of automated stuttered speech, researchers have explored the transformer-based model Wav2vec2.0 which has shown commendable performances in various studies \cite{mohapatra2022speech}, \cite{bayerl22_interspeech}, \cite{bayerl2022ksof}, \cite{al2024novel}. The labeled datasets SEP-28k and FluencyBank proposed by \cite{lea2021sep}, has been acquired by the researchers. However, the majority of the studies are on single disfluency types, despite the fact that multiple disfluencies can occur in a single audio clip as well. In this section, we will cover the studies addressing multi-stutter labels. 

Disfluencies often co-occur with other types of disfluencies in real-world scenarios. Mapping this issue into deep learning parlance, each instance may consist of multiple labels \cite{tsoumakas2007multi} , framing this as a multi-label classification problem. In the domain of speech processing, numerous researchers have employed multi-label classification for emotion recognition \cite{slimi2022multiple}, voice disorders \cite{junior2023multiple} and speech recognition use-cases \cite{hossain2022multi}. 
In the context of multi-stuttered scenarios, the authors \cite{jouaiti2022multi} have conceptualized the issue as a multi-label classification problem. They have proposed ResNet-50 as a solution for converting audio into spectrograms and feeding the images in the network. The researchers have utilized SEP-28k and FluencyBank datasets for training and testing. While the paper acknowledges the issue regarding data imbalance, it doesn't explicitly address the resolution of the problem.  In addition, the splits of the datasets play a major role in the outcome of the model, however, they are not discussed in the paper. The influence of dataset splits has been explored in the study \cite{bayerl2022influence}, as it highlights the capability of the model’s efficacy.

The authors \cite{bayerl2023stutter} and \cite{wagner2024large} have also taken it as a multi-label problem and explored the cross-language capabilities of a model so it can be more generalized. In contrast to the aforementioned study \cite{jouaiti2022multi}, this study explicitly mentions the training, validation, and testing splits of the datasets. This study uses the SEP-28k Extended dataset, and KSoF dataset which is based on English and German stuttered speech. The proposed method uses Wav2Vec2.0 along with a sequence classifier for end-to-end stuttering detection and classification. Furthermore, they have applied multi task learning to incorporate regularization. The evaluation metric acquired was F1-score for different combinations of training data and modifications. For the SEP-28k-E test set the F1-scores of Block, Interjection, Prolongation, Sound Repetition, and Word Repetition achieved were 0.32, 0.77, 0.54, 0.50, and 0.56, respectively. Furthermore, for the FluencyBank dataset, the F1-scores were 0.31, 0.82, 0.57, 0.61 and 0.45 in correspondence with the above mentioned disfluency types. This has provided a new direction for solving the task at hand, however, the paper does not cater to the quality of the data by omitting the discussions on inter-annotator agreements in the labeling of these disfluencies. In Contrast, the study proposed by \cite{mohapatra2022speech} has produced better results by only acquiring the high inter-annotator agreement labels. They have only considered the instances where all three annotators agreed on the labels. So, even if the quantity of the data was decreased, better results were achieved due to the quality of the dataset. In addition, they have neither applied any preprocessing steps to the audio clips nor catered multi-stuttered use-case scenario. 

The significance of preprocessing has been established in the paper \cite{ameer2023whisper}. The study has followed a series of data cleaning and preprocessing steps with significant improvement in the results. Furthermore, the approach proposed by \cite{mohapatra2022speech} regarding the data quality of quantity was also used in this paper.  The dataset used in this study is SEP-28k for training and FluencyBank for testing purposes. The paper acquires Whisper for the classification of disfluencies in stuttered speech, demonstrating superior performance compared to results obtained with the Wav2Vec2.0 model. The paper has experimented with various data-splits achieving the highest average F1-score of 0.81. However, the study does not consider the multi stuttering cases. 

Hence, there exists a notable gap in the literature which necessitates the integration of multi-label classification and considers the proper dataset splits of training, validation and testing sets.  Next, with the inclusion of preprocessing and cleaning of the input data based on transformer-based models, this research can be taken forward for real-world scenarios. However, another important aspect that needs to be addressed is the multi-label data for stuttered speech. While SEP-28k and FluencyBank encompass certain occurrences of multi-label instances. On the other hand, by prioritizing the dataset quality and choosing to include only instances where all three annotators agree on the same labels, the multi-label number of instances was reduced significantly. In this context, scarcity presents a challenge to the learning of the models. Therefore, this emphasizes the need for a dataset that specifically is tailored to multi-stuttered speech. Lastly, building upon the findings presented in the referenced paper \cite{ameer2023whisper}, where Whisper has outperformed Wav2Vec2.0. Our study, in conjunction with other mentioned gaps, will also leverage Whisper for the automated classification of disfluencies in the context of multi-stuttered speech.

\section{Methodology}
\label{sec:methodology}
This study addresses the crucial gaps identified in the existing literature concerning the classification of disfluencies in multi-stuttered instances. In this section, we have first discussed the details related to datasets along with the curation of multi-stuttered data. Next, we comprehensively discussed the architecture of Whisper, followed by the experimental design employed in this study and the corresponding experimental setup.
\subsection{Datasets}
\label{subsec:datasets}
In this study, two primary datasets SEP-28k \cite{lea2021sep} and FluencyBank \cite{ratner2018fluency} have been leveraged for training and testing the models. SEP-28k consists of ~28k instances, and each audio clip is 3 seconds long. These clips were extracted from eight podcasts and labeled by experts with five major disfluency types; prolongation, sound repetition, word repetition, interjections, and blocks \cite{lea2021sep}. Similarly, in the paper \cite{lea2021sep}, FluencyBank has also been labelled and consists of the same audio length as SEP-28k. It consists of ~4k instances and is mostly leveraged for testing purposes. Lastly, both datasets consist of audio of English speakers. In addition, in this study, SEP-28k have been used for training, validation and internal testing. FluencyBank has been used for the external testing of the model.
In the study presented by \cite{bayerl2022influence}, the influence of data partitioning has been explored. The authors have proposed and experimented with three speaker-exclusive dataset splits i.e. SEP-28k-E, SEP-28k-E-T and SEP-28k-D. The training split of SEP-28k-E consists of clips with the top four dominant speakers. The remaining clips are partitioned for testing and validation. SEP-28k-T and SEP-28k-D are similar as they are using the testing and validation split of SEP-28k-E in their training splits, and testing them on the top four dominant speakers. The splits of SEP-28k-E can exhibit the capability of a model by training them on a few speakers with more audio clips and testing on various speakers. On the other hand, SEP-28k-T and SEP-28k-D can identify the ability of the model to learn from fewer examples of various speakers. In the experimental work of \cite{ameer2023whisper}, the authors have investigated a different setup that combines training and validation data, using the test split for validation purposes. These datasets, which we will refer to as SEP-28k-E-merged and SEP-28k-T-merged, are shown in Table \ref{tab:data-split-abbrev} with their splits. These additional splits are also included in the respective study.

To improve the dataset's quality, data cleaning steps from the paper \cite{ameer2023whisper} have been followed considering their improved outcomes. At first, the experimentation was limited to cases where all three annotators agreed on a certain label. Next, labels that accounted for less than 1\% of the dataset's representation were eliminated. These were non-disfluent labels, like Natural Pause, Hard to Understand, Speechless, Bad Audio Quality, and Music. Moreover, in order to maintain uniformity, audio fragments that lasted less than three seconds were also not included. To reduce the possibility of influencing the outcome of the model, background noises were reduced from the audio clips. Finally, audio samples that included multiple speakers were also excluded as they may complicate classification and interpretation.

\begin{table}[htbp]
\centering
\caption{Split Configurations for Experimental Study}
\label{tab:data-split-abbrev}
\renewcommand{\arraystretch}{1.2} % Adjust spacing between rows
\begin{adjustbox}{max width=\textwidth}
\begin{tabular}{@{}lll l@{}}
\toprule
Data Split & Training Data & Validation Data & Testing Data \\
\hline
SEP-28k-E & 4-DS & DS-Set 1 & DS-Set 2 \\
SEP-28k-T & DS-Set 1 & DS-Set 2 & 4-DS \\
SEP-28k-D & DS-Set 2 & DS-Set 1 & 4-DS \\
SEP-28k-E-merged & 4-DS + DS-Set 1 & DS-Set 2 & FB \\
SEP-28k-T-merged & DS-Set 1 + DS-Set 2 & 4-DS & FB \\
\bottomrule
\end{tabular}
\end{adjustbox}
%\vspace{0.01in}
\par \footnotesize
Note -- DS = Dominant speakers, DS-Set = Distinct speakers-Set, FB = FluencyBank 
\par \normalsize
\end{table}

\subsection{Data Curation for Multi-Stutter}
\label{sec:data curation}
In the course of implementing data cleaning and preprocessing procedures, it was observed that the resultant dataset had few instances with multi-stuttered labels, their count is demonstrated in Table \ref{tab:combination_counts}. Given the insufficient count and limited combinations of multi-stuttered instances, there is a compelling need for a multi-stuttered dataset.

\begin{figure}[h]
    \centering
    % Option 1: Using the `svg` package
  %  \includesvg[width=\linewidth]{whisper.svg}
    \includegraphics[width=\linewidth]{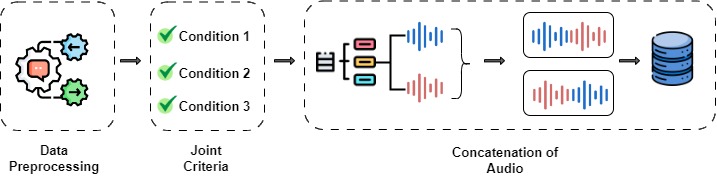}
  
    \caption{Pipeline for Curation of Multi-stuttered data: Condition 1 involves acquiring audio from the same episodes, Condition 2 requires audios featuring the same speakers, and Condition 3, proceeds with concatenation if two audios have distinct disfluency labels or both have the label "NoStutteredWords".}
    \label{fig:audio concatenation}
\end{figure}

\begin{table}[h]
    \centering
    \caption{Count of Instances with Multi-Stuttered Labels in the Preprocessed Dataset}
    \label{tab:combination_counts}
    \begin{tabular}{lc}
        \toprule
        Combination & Count \\
        \midrule
        WordRep\_Interjection\_ & 178 \\
        Prolongation\_Interjection\_ & 90 \\
        SoundRep\_Interjection\_ & 49 \\
        Block\_Interjection\_ & 24 \\
        SoundRep\_WordRep\_ & 16 \\
        Prolongation\_WordRep\_ & 11 \\
        Block\_SoundRep\_ & 8 \\
        Block\_WordRep\_ & 6 \\
        Prolongation\_SoundRep\_ & 5 \\
        Prolongation\_Block\_ & 4 \\
        SoundRep\_WordRep\_Interjection\_ & 3 \\
        Block\_SoundRep\_Interjection\_ & 1 \\
        \midrule
        Total & 395 \\
        \bottomrule
    \end{tabular}
\end{table}
Therefore, to curate multi-stuttered data, specific conditions needed to be met. The audio from the same episodes and the same speakers were required. Since in each episode featured multiple speakers, we manually had to label each of them so they could be concatenated. Next, two audio samples consisting of distinct disfluent labels were then concatenated using Python library AudioSegment. In addition, if both the labels were "NoStutteredWords", they were also concatenated. The pipeline for the multi-stuttered dataset is illustrated in Figure \ref{fig:audio concatenation}.

After the concatenation, the audio clips are 6 seconds with a 16 kHz sampling rate. Furthermore, to ensure label balance, the NoStutteredWords category was addressed by sampling from each speaker and subsequently calculating the average for this particular label. The Multi-Stuttered dataset was partitioned based on the splits mentioned in Table \ref{tab:data-split-abbrev}, the DS have 47951 instances, DS-Set 1 have 11857 instances, and DS-Set 2 have 10512 instances, and FB have 11806 instances, the details of the counts and combinations are presented in the Table \ref{tab:multi-stuttered-combination_counts}
\begin{table}[ht]
    \centering
    \caption{Distribution of Instances in the Multi-Stuttered Dataset Across Different Splits, Including Total Counts and Specific Allocations for DS-Set 1, DS-Set 2, DS.}
    \begin{tabular}{lrrr}
        \toprule
        Multi-stuttered Combinations & DS-Set 1 & DS-Set 2 & DS \\
        \midrule
        SoundRep\_Interjection\_ & 1482 & 559 & 3033 \\
        Interjection\_SoundRep\_ & 1482 & 559 & 3033 \\
        Interjection\_WordRep\_ & 974 & 1741 & 7637 \\
        WordRep\_Interjection\_ & 974 & 1741 & 7637 \\
        NoStutteredWords\_NoStutteredWords\_ & 915 & 1062 & 3817 \\
        SoundRep\_WordRep\_ & 561 & 395 & 1989 \\
        WordRep\_SoundRep\_ & 561 & 395 & 1989 \\
        Interjection\_Prolongation\_ & 535 & 502 & 2589 \\
        Prolongation\_Interjection\_ & 535 & 502 & 2589 \\
        SoundRep\_Prolongation\_ & 427 & 173 & 997 \\
        Prolongation\_SoundRep\_ & 427 & 173 & 997 \\
        Block\_Interjection\_ & 351 & 435 & 1720 \\
        Interjection\_Block\_ & 351 & 435 & 1720 \\
        Prolongation\_WordRep\_ & 331 & 354 & 1276 \\
        WordRep\_Prolongation\_ & 331 & 354 & 1276 \\
        WordRep\_Block\_ & 328 & 253 & 1251 \\
        Block\_WordRep\_ & 328 & 253 & 1251 \\
        Block\_SoundRep\_ & 281 & 129 & 881 \\
        SoundRep\_Block\_ & 281 & 129 & 881 \\
        Block\_Prolongation\_ & 201 & 184 & 694 \\
        Prolongation\_Block\_ & 201 & 184 & 694 \\
        \bottomrule
    \end{tabular}
    \label{tab:multi-stuttered-combination_counts}
\end{table}

\subsection{Model Architecture}
Whisper is an encoder-decoder architecture based on weakly supervised transformers that are state-of-the-art at the moment, particularly for speech recognition \cite{radford2023robust}. Whisper uses a 16 kHz sampling rate for the audio clips that are fed into it. Next, these audio clips are transformed into a log Mel spectrogram representation. After that, two 1-D convolutional layers are used to process them in order to extract features. Then, depending on the requirements of a given task, these features are fed into an encoder and a decoder. To enable the model to be used for a variety of use cases, such as speech recognition, language identification, voice detection, and speech translation, the decoder is trained in a multitask format. However, for the classification of multi-stuttered disfluency types, only the encoder part of the model is required. 

\subsubsection{Feature Extractor}
\label{subsubsec:feature extractor}
Whisper uses the log Mel spectrogram, or more precisely, the log-magnitude Mel spectrogram, which is represented by the symbol \( S_{\text{log-Mel}} \). First, using a window function \( w(t) \) and represented as \( x(t) \), the audio signal is windowed and split into short frames: 
\[ x_w(t) = x(t) \cdot w(t) \]
The signal is then transformed from the time domain to the frequency domain for each windowed segment using the Fast Fourier Transform (FFT), resulting in a complex-valued spectrogram \( X(f, \tau) \): 
\[ X(f, \tau) = \text{FFT}[x_w(t)] \] 
%The formula is \( W_{\text{length}} = 25 \, \text{ms} \) and \( %W_{\text{stride}} = 10 \, \text{ms} \).
The spectrogram is then passed through the Mel filterbank, represented by \( M(f) \), which uses overlapping triangle filters on the Mel scale to extract energy in particular frequency ranges:
\[ X(f, \tau) = \sum_{f} X(f, \tau) \cdot M_m(f) = Y_m(\tau) \]
The summed energy values \( Y_m(\tau) \) from each filterbank output are logarithmically transformed using the natural logarithm, represented as \( \log(\cdot) \), which lowers the data's dynamic range and increases perceptual relevance:
\[ S_{\text{log-Mel}}(m, \tau) = \log(Y_m(\tau)) \]
The resulting log Mel spectrogram, which is useful for tasks like audio classification, is a two-dimensional matrix \( S_{\text{log-Mel}} \). It captures important spectral features over time. This log Mel spectrogram is calculated by the Whisper feature extractor using 25-millisecond windows and a 10-millisecond stride.

\subsubsection{Feature Normalization}
\label{subsubsec:feature normalization} 
The input features extracted from log Mel spectrograms are subjected to feature normalization as a preprocessing step. This step scales the features globally to fall between -1 and 1, ensuring uniformity and optimal model training. The training process is more stable when the pre-training dataset has a mean that is almost zero.

\subsubsection{Preliminary Convolutional Processing}
\label{subsubsec:preliminary convolutional processing}
After the normalization, two convolution layers form a small stem applied to the input representation.  Filter widths for both convolution layers are 3, and for the second convolution layer stride is kept 2. After every convolutional layer, the Gaussian Error Linear Unit (GELU) activation function is also applied which adds non-linearity to the network.

\subsubsection{Sinusoidal Positional Embedding}
\label{subsubsec:sinusoidal positional embedding} 
To improve the model's capacity to extract positional relationships and sequential information from the input representation, the stem's output is added with sinusoidal position embeddings. The following formula can be used to create these embeddings:
The expression \[
\text{Position\_Embed}(pos, 2i) = \sin\left(\frac{pos}{10000^{(2i / d)}}\right)
\]
\[
\text{Position\_Embed}(pos, 2i+1) = \cos\left(\frac{pos}{10000^{(2i / d)}}\right)
\]
In this case, the sine component for the \(i\)-th dimension is represented by \( \text{Position\_Embed}(pos, 2i) \) and the cosine component by \( \text{Position\_Embed}(pos, 2i+1) \). The position of the token in the sequence is indicated by the term \(pos\), the dimension index is indicated by \(i\), and the total dimensionality of the positional embeddings is indicated by \(d\). Since each dimension captures distinct periodicities due to the exponential increase in frequency across dimensions, order and temporal information within the sequence are effectively encoded.
  
\subsubsection{Encoder layer of Transformer}
\label{subsubsec:encoder layer of transformer} 
The sinusoidal position embeddings are then fed to encoder Transformer blocks. Typically, feedforward neural networks and self-attention layers make up each encoder block.  The model can capture dependencies between elements by adaptively focusing on different parts of the input sequence thanks to self-attention mechanisms. 

The self-attention mechanism allows elements of the input sequence to selectively focus on different segments; the Query, Key, and Value matrices are represented by the symbols \(Q\), \(K\), and \(V\). The softmax operation is used to normalize the attention scores, resulting in weighted sums that produce attention-weighted representations.
\begin{equation}
      \text{Attention}(Q, K, V) = \text{softmax}\left(\frac{QK^T}{\sqrt{d_k}}\right) V
   \end{equation}
The feedforward network, which comes after the self-attention mechanism, independently transforms each position in a non-linear way. 
\begin{equation}
\text{FFN}(x) = \max(0, xW_1 + b_1)W_2 + b_2
\end{equation}
Following each sub-layer, layer normalization and residual connections are applied.
   \begin{equation}
      \text{LayerNorm}(x) = \frac{x - \mu}{\sigma} \odot \gamma + \beta
   \end{equation}
   where \(\mu\) and \(\sigma\) represent the mean and standard deviation of \(x\), and \(\gamma\) and \(\beta\) are learnable parameters.
The output of each sub-layer is given by:
   \begin{equation}
      \text{SubLayer}(x) = \text{LayerNorm}(x + \text{SubLayer\_Output}(x))
   \end{equation}
Multiple Transformer encoder layers are sequentially stacked to create a deep hierarchical architecture. The output of one layer serves as the input to the subsequent layer.
\subsubsection{Dimensionality Reduction: Projector Layer}
\label{subsubsec:projector layer} 
The learned representations' dimensionality from the stacked encoder is decreased by the projector, a linear layer. The model in question projects the 512-dimensional output of the WhisperEncoder into a 256-dimensional vector, which is a lower-dimensional space. In order to reduce computational complexity and extract the most important features from the input data, dimensionality reduction is frequently used. Before the last classification stage, the projector essentially transforms features.

\subsubsection{Linear Classification Layer}
\label{subsubsec:linear classification layer} 
The classifier is an additional linear layer that performs the final classification using the projector's output, which is now a 256-dimensional vector. In this instance, the number of classes in the classification task is reflected in the output space, which is mapped by the classifier from the 256-dimensional representation. This model has been created for a six-output class classification task.
\subsubsection{Loss Function}
\label{subsubsec: loss function}

Loss functions play a significant role in the evaluation of a model as it quantifies between predictions and actual results. Since our problem is a multi-label classification of multiple disfluences, the Binary Cross-Entropy (BCE) loss function is well-suited \cite{bayerl2023stutter}. More specifically, we have used BCEWithLogitsLoss function in Pytorch 
\footnote{\url{https://pytorch.org/docs/stable/generated/torch.nn.BCEWithLogitsLoss.html}}, it combines sigmoid function and BCE Loss in a single operation. Therefore, it can be acquired for instances which have multiple labels. The mathematical formula for BCE Loss is as follows
\[
\text{BCE Loss} = - \frac{1}{N} \sum_{i=1}^{N} \left[ y_i \cdot \log(p_i) + (1 - y_i) \cdot \log(1 - p_i) \right]
\]
where:
\begin{itemize}
    \item \(N\) is the number of instances in the dataset.
    \item \(y_i\) is the actual binary label for the \(i\)-th example (0 or 1).
    \item \(p_i\) is the predicted probability for the \(i\)-th example belonging to the positive class.
\end{itemize} 

\subsection{Experimental Design}
\label{subsec:experimental design}
After curating the dataset for multi-stuttered speech, speaker-exclusive data splits discussed in the section \ref{subsec:datasets} are then fine-tuned. In this study, to cater for multi-stuttered instances, encoder-only Whisper has been leveraged by fine-tuning its pre-trained English base model. In order to adapt the model to our specific task, we used Hugging Face Library to acquire its module namely, Whisper for Audio Classification \footnote{\url{https://huggingface.co/docs/transformers/model_doc/Whisper}}. 
\begin{figure}[h]
    \centering
    % Option 1: Using the `svg` package
  %  \includesvg[width=\linewidth]{whisper.svg}
    \includegraphics[width=\linewidth]{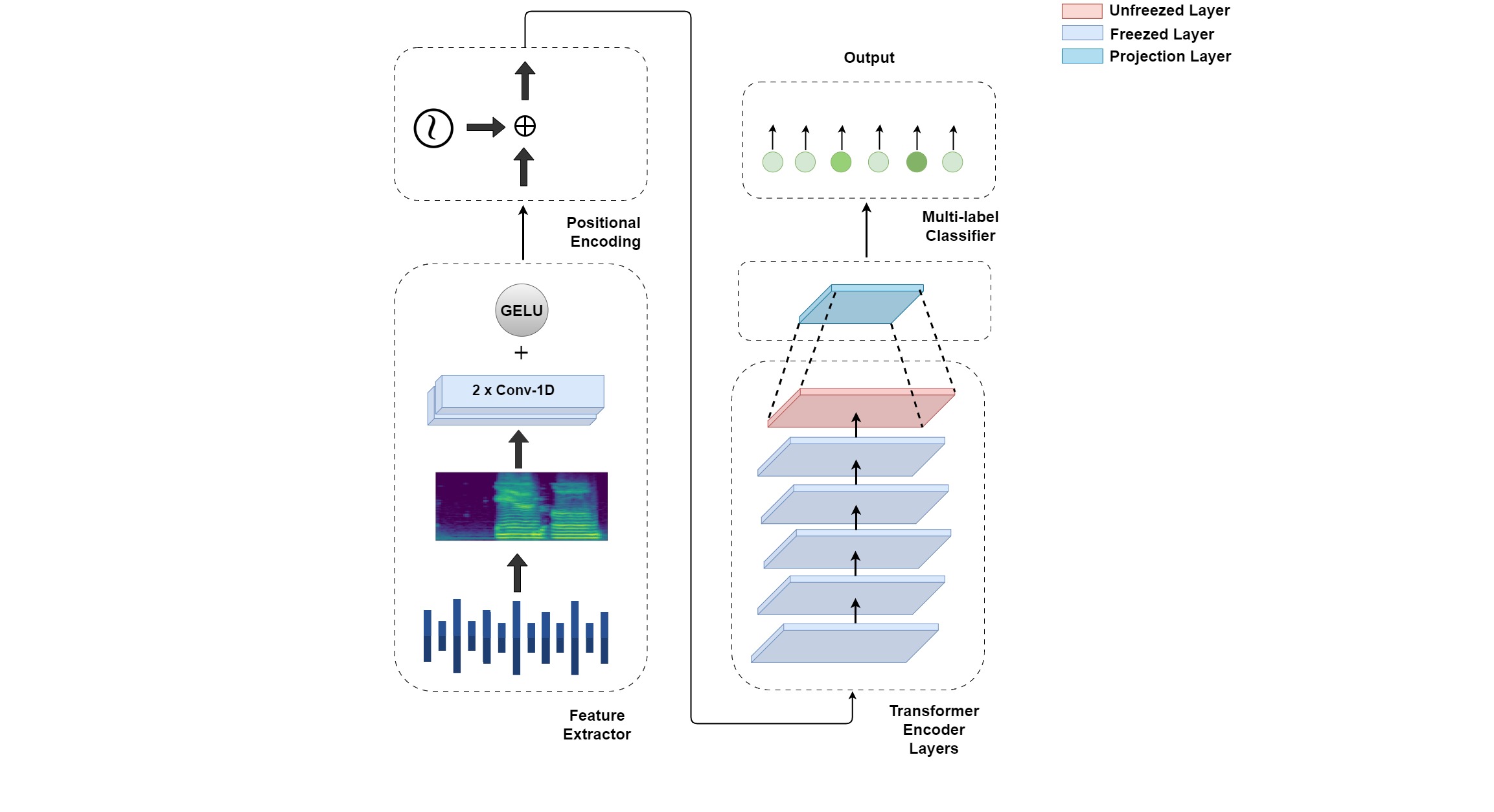}
  
    \caption{Whisper Architecture Configuration Designed for Enhanced Classification of Multi-Stuttered Speech Patterns}
    \label{fig:proposed architecture settings multi}
\end{figure}
First, the base models of Wav2Vec2.0 and Whisper were finetuned on the multi-stuttered dataset on all the splits. Based on the results, Whisper was chosen for further experimentation. Next, the encoder layer optimization strategy was applied, in which the first three encoder layers of the Whisper model were frozen while fine-tuning. The insights of the optimization strategy were gained from earlier research \cite{ameer2023whisper}.  For further reduction of trainable parameters, we have also experimented with the freezing of the feature extractor. In addition, we have also included the experimentation of freezing the first four, first five layers, and all layers of the encoder. Lastly, the Whisper model with only four encoder layers was also fine-tuned. However, the results exhibited a significant decline, which led us to discontinue further experimentation. Detailed outcomes stemming from this experimental design are presented and discussed in the section \ref{sec:results and discussion}.

\subsection{Experimental Setup}
\label{subsec:experimental Setup}

The details of the parameters and other specifications can be seen in Table 
\ref{tab:model_params}. It is worth noting that the chosen evaluation metric was F1-score as it takes both precision and recall into account and has been a preferred metric in previous literature as well. Also, to avoid model over-fitting, early stopping was enabled.  

\begin{table}[htbp]
\centering

\caption{Model Fine-Tuning Specifications}
\label{tab:model_params}
\renewcommand{\arraystretch}{1.2} % Adjust spacing between rows
\begin{adjustbox}{max width=\textwidth}
\begin{tabular}{@{}lc@{}}
\toprule
\textbf{Configuration} & \textbf{Details} \\
\midrule
GPU & NVIDIA GeForce RTX 2070 \\
Batch Size & 8 \\
Pre-Trained Model & whisper-base.en \footnote{\url{https://huggingface.co/openai/whisper-base.en}}
\\
Loss Function & Binary Cross Entropy \footnote{\url{https://pytorch.org/docs/stable/generated/torch.nn.BCEWithLogitsLoss.html}}\\
Learning Rate & \num{2.5e-5} \footnote{\url{https://github.com/vasistalodagala/whisper-finetune}} \\
Chunk Length & 6.0  \\
Sampling Rate & 16 kHz \\
Evaluation Metric & F1-score \\
Early Stopping & Enabled \\
\bottomrule
\end{tabular}
\end{adjustbox}
\end{table}

\section{Results and Discussion}
\label{sec:results and discussion}
In this section, we begin by analyzing the outcomes between the Wav2Vec2.0 and Whisper models. The results led us to continue with the Whisper model for further experimentation to achieve the classification of multi-stuttered disfluencies in speech. Next, we have explored the impact of freezing the feature extractor on the outcome. The findings demonstrated that freezing the feature extractor does not affect the results, thereby preferring the respective configuration with reduced trainable parameters. Finally, building on the paper \cite{ameer2023whisper}, we extended our experimentation to the remaining encoder layers. Through these experiments, we identified the optimal layer freezing configuration of the model, and compared with with \cite{bayerl2023stutter}. 

In the literature, the research on multi-stuttered classification of disfluency is notably scarce. In the study  \cite{bayerl2023stutter}, authors have focused on this problem leveraging Wav2Vec2.0 for the classification of disfluency types in multi-stuttered speech. However, in the paper, \cite{ameer2023whisper} Whisper appeared to be a more suitable choice for this task. To make a concrete choice between these two models we fine-tuned both Wav2Vec2.0 and Whisper. The comparative analysis in Table \ref{tab:wav2vec2.0vswhisper} demonstrates that Whisper is a better choice across all the data splits in contrast to Wav2Vec2.0. For this experimentation, Wav2Vec2.0’s base model which consists of 12 encoder layers was acquired. In contrast, Whisper’s base model consisting of 6 encoder layers was fine-tuned. All the Layers were kept unfrozen in both of the models. It is noticeable in the Table \ref{tab:wav2vec2.0vswhisper} that there is a considerable variation of results between the models. Given their architectures, one of the key differences between Wav2Vec2.0 and Whisper is their approaches to input representation of waveforms, potentially leading to substantial differences in results. Whisper converts the raw input into mel-spectrograms which are then passed to convolutional layers for feature extraction. On the other hand, Wav2Vec2.0 directly extracts these features from raw audio through convolutional layers. Applying the convolutional layers directly on raw waveforms extracts more features as they are learned through trainable parameters. It seems that acquiring features through convolutional layers should have better outcomes as it is learning more features but it may not be necessary. For the respective task, features that are related to human speech are more relevant and also limit the computational head on the models. Therefore, one of the major reasons Whisper is performing better could be due to the usage of mel-spectrograms for feature extraction. Mel-spectrograms mimic the human auditory system, and these finite features are then passed on to convolutional layers and eventually to the encoder layers. 

\begin{table}[htbp]
\centering
\caption{Comparative Performance of Wav2Vec2.0 and Whisper Models on FluencyBank Dataset}
\label{tab:wav2vec2.0vswhisper}
\begin{tabularx}{\textwidth}{@{}lXXXXXXX@{}} % Added one more X for the spacer column
\toprule
\textbf{Dataset} & \multicolumn{3}{c}{\textbf{Wav2vec2.0}} & \textbf{} & \multicolumn{3}{c}{\textbf{Whisper}} \\
\cmidrule(r){2-4} \cmidrule(l){6-8} % Adjusted for the spacer column
& \textbf{micro} & \textbf{macro} & \textbf{weighted} & & \textbf{micro} & \textbf{macro} & \textbf{weighted} \\
\midrule
SEP28k-E & 0.44 & 0.13 & 0.26 & & 0.76 & 0.71 & 0.75 \\
SEP28k-T & 0.55 & 0.23 & 0.40 & & 0.73 & 0.66 & 0.71 \\
SEP28k-D & 0.44 & 0.13 & 0.26 & & 0.76 & 0.71 & 0.75 \\
SEP28k-E-Merged & 0.44 & 0.13 & 0.26 & & 0.73 & 0.67 & 0.71 \\
SEP28k-T-Merged & 0.44 & 0.13 & 0.26 & & 0.76 & 0.72 & 0.75 \\
\bottomrule
\end{tabularx}
\vspace{0.5mm} % Adjust the space as needed, here it's set to 5mm
\footnotesize Note: The terms micro, macro, and weighted refer to the types of F1-scores.
\end{table}

\begin{table}[htbp]
\centering
\caption{Comparative Analysis of Model Performance with Frozen and Unfrozen Feature Extractor Across SEP-28k-E and FluencyBank Datasets}
\label{tab:feature extractor comparison}
\begin{tabular}{lcccc}
\toprule
\multicolumn{5}{c}{\textbf{SEP-28k-E}} \\ % Centered header for SEP-28k-E
\midrule
\textbf{} & \textbf{UnFrz0-5} & \textbf{UnFrz0-5+FrzFE} & \textbf{Frz0-2} & \textbf{Frz0-2+FrzFE} \\
\midrule

Micro & 0.83 & 0.82 & 0.84 & 0.85 \\
Macro & 0.79 & 0.78 & 0.81 & 0.83 \\
Weighted & 0.82 & 0.81 & 0.83 & 0.85 \\
\midrule

Block  & 0.50 & 0.38 & 0.46 & 0.50 \\
Interjection  & 0.83 & 0.83 & 0.88 & 0.88 \\
Prolongation & 0.72 & 0.70 & 0.62 & 0.68 \\
Sound Repetition & 0.69 & 0.69 & 0.62 & 0.71 \\
Word Repetition & 0.75 & 0.76 & 0.54 & 0.86 \\
No Stutter Words & 0.68 & 0.57 & 0.70 & 0.78 \\
\midrule
\multicolumn{5}{c}{\rule{0pt}{3ex}\textbf{FluencyBank}} \\ % Centered header for FluencyBank
\midrule

Micro & 0.76 & 0.75 & 0.79 & 0.80 \\
Macro & 0.71 & 0.70 & 0.75 & 0.76 \\
Weighted & 0.75 & 0.74 & 0.78 & 0.79 \\
\midrule

Block  & 0.38 & 0.30 & 0.44 & 0.45 \\
Interjection & 0.79 & 0.79 & 0.81 & 0.81 \\
Prolongation  & 0.36 & 0.57 & 0.59 & 0.60 \\
Sound Repetition & 0.58 & 0.54 & 0.71 & 0.72 \\
Word Repetition & 0.59 & 0.63 & 0.69 & 0.70 \\
No Stutter Words & 0.60 & 0.31 & 0.40 & 0.45 \\
\bottomrule
\end{tabular}

\footnotesize 
 Note: In the table, "UnFrz0-5" denotes layers 0 to 5 are unfrozen (trainable), "UnFrz0-5+FrzFE" indicates layers 0 to 5 are unfrozen with the feature extractor (FE) being frozen, "Frz0-2" signifies only layers 0, 1, and 2 are frozen (not trainable), and "Frz0-2+FrzFE" means layers 0, 1, and 2 are frozen along with the feature extractor.
\normalsize
\end{table}

Since it has been established that Whisper is the better choice for the task at hand. Based on the groundwork laid by \cite{ameer2023whisper}, it was proved through experimentation that freezing the initial layers of the encoder model will yield better results. Consequently, this methodology will have fewer trainable parameters. Therefore, we experimented with this particular configuration of freezing the first three layers of the model. Building on this study, we also experimented with freezing the feature extractor for further reduction of parameters. The results presented in the Table \ref{tab:feature extractor comparison} demonstrate that the outcomes of the freezed feature extractor are comparable to the unfreezed feature extractor. Thus, the rest of the experiments are done keeping the feature extractor layers frozen. Table \ref{tab:parameters} shows the comparison between the reduced learnable parameters.

\begin{table}[htbp]
\centering
\caption{Model Configuration Parameters in Millions}
\label{tab:parameters}
\begin{tabular}{lc}
\toprule
\textbf{Configuration} & \textbf{Parameters (Millions)} \\
\midrule
UnFrz0-5 & 20.72 \\
\textbf{UnFrz0-5+FrzFE} & \textbf{19.05} \\
Frz0-2 & 11.27 \\
\textbf{Frz0-2+FrzFE} & \textbf{9.59} \\
\bottomrule
\end{tabular}
\end{table}
\begin{table}[htbp]
\centering
\caption{Model Performance with Various Layer Freezing Configurations of Best Data split}
\label{tab:results}
\footnotesize % Reduce font size to fit the table better on the page
\begin{tabularx}{\linewidth}{l *{5}{>{\centering\arraybackslash}X}}
\toprule
\textbf{Metric} & \textbf{UnFrz0-5+FrzFE} & \textbf{Frz0-2+FrzFE} & \textbf{Frz0-3+FrzFE} & \textbf{Frz0-4+FrzFE} & \textbf{Frz0-5+FrzFE} \\
\midrule
\multicolumn{6}{c}{\textbf{SEP-28k-E}} \\
\midrule
Micro & 0.82 & 0.85 & 0.85 & 0.88 & 0.73 \\
Macro & 0.78 & 0.83 & 0.85 & 0.85 & 0.68 \\
Weighted & 0.81 & 0.85 & 0.85 & 0.87 & 0.73 \\
\midrule
Block & 0.38 & 0.5 & 0.48 & 0.41 & 0 \\
Interjection & 0.83 & 0.88 & 0.88 & 0.91 & 0.76 \\
Prolongation & 0.70 & 0.68 & 0.73 & 0.74 & 0 \\
SoundRep & 0.69 & 0.71 & 0.64 & 0.67 & 0 \\
WordRep & 0.76 & 0.86 & 0.87 & 0.9 & 0.68 \\
No Stutter Words & 0.57 & 0.78 & 0.73 & 0.83 & 0 \\
\midrule
\multicolumn{6}{c}{\textbf{FluencyBank Results}} \\
\midrule
Micro & 0.75 & 0.8 & 0.78 & 0.82 & 0.71 \\
Macro & 0.70 & 0.76 & 0.74 & 0.78 & 0.67 \\
Weighted & 0.74 & 0.79 & 0.78 & 0.81 & 0.71 \\
\midrule
Block & 0.30 & 0.45 & 0.43 & 0.34 & 0 \\
Interjection & 0.79 & 0.81 & 0.8 & 0.83 & 0.79 \\
Prolongation & 0.57 & 0.6 & 0.61 & 0.56 & 0 \\
SoundRep & 0.54 & 0.72 & 0.66 & 0.69 & 0 \\
WordRep & 0.63 & 0.7 & 0.69 & 0.73 & 0.69 \\
No Stutter Words & 0.31 & 0.45 & 0.4 & 0.54 & 0 \\
\bottomrule
\end{tabularx}
\end{table}
Keeping this configuration of the frozen feature extractor, we also froze the remaining three layers one by one and analyzed the results. Table \ref{tab:results} presents the average and class-wise F1-scores of each frozen layer strategy. Notably, it can be observed that the F1-scores of the first three layers frozen and the first five layers frozen are comparable. Analyzing the results of freezing the first five layers and all six layers, a significant difference is apparent in Table \ref{tab:results}. Hence, this observation highlights the contribution of the last layer in the identification of multi-stuttered disfluencies. Therefore, freezing the top 5 layers is the most optimal configuration for the respective task reducing the learnable parameters to 3.29 million. 

\begin{figure}[h]
    \centering
    % Option 1: Using the `svg` package
  %  \includesvg[width=\linewidth]{whisper.svg}
    \includegraphics[width=\linewidth]{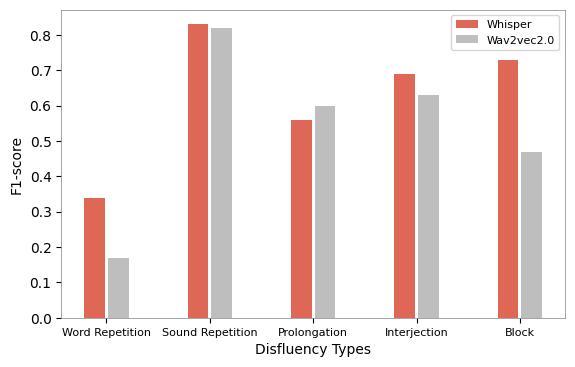}
  
    \caption{Comparative Performance Analysis Showing Superior Results of Whisper Over Wav2vec2.0 with Fewer Trainable Parameters}
    \label{fig:whisper vs wav2vec2.0}
\end{figure}

We have also compared our results from the top five frozen layers with the outcomes reported by \cite{bayerl2023stutter}. The Figure \ref{fig:whisper vs wav2vec2.0} illustrated that our model outcomes have outperformed the Wav2Vec2.0 proposed by \cite{bayerl2023stutter} with considerably less trainable parameters. In addition, it’s noteworthy that the Whisper model’s variant with 6 encoder layers was used in our experiments, and Wav2Vec2.0 model’s large variant which consists of 24 encoder layers was used for the respective task. Therefore, Wav2Vec2.0 large variant had 315.70 million trainable parameters, and our proposed whisper model configuration had 3.29 million trainable parameters.

Thus, through our experimental design, we have reached the conclusion that the last encoder layer of Whisper has a significant effect as it plays an eminent role in the identification of multi-stuttered disfluency types. In addition, we have fine-tuned the model with fewer trainable parameters without compromising the results. Our experimentation is a step forward towards a resource efficient approach for this task, as we were able to narrow down to the most contributing layer. The proposed model configurations are presented in Figure \ref{fig:proposed architecture settings multi}. Lastly, the research area is still open for the classification of stuttered speech with respect to multilingual use cases.
\subsection{Theoretical and Practical Implications}
This study validates the effectiveness of Whisper in multi-stuttered speech classification. The exploration of layer freezing strategies has revealed insights into the features that contribute most to stuttered speech classification. It has been identified through experimentation in our study that freezing the first five layers yields the most optimal results. The comparative analysis with the state-of-the-art Wav2Vec2.0 implies the superiority of Whisper in terms of both trainable parameters and F1-scores. In addition, the Whisper's approach to acquire mel-spectrograms for feature extraction could  contributing significantly contribute to better performance than Wav2Vec2.0, as it mimics human auditory system.

Since the proposed approach has outperformed Wav2Vec2.0 with significantly fewer trainable parameters, it suggests the potential for an automated assessment system to aid speech-language pathologists in diagnosis. The resource-efficient nature of this study enhances its feasibility for real-world applications. Researchers and developers can leverage these findings to improve results and expand into different languages.

\section{Conclusion}
\label{sec:conclusion}
This study aims to classify multi-stuttered disfluencies, a research area seldom explored in the community. By curation of a multi-stuttered dataset, and through extensive experimentation, we were able to achieve notable outcomes by acquiring Whisper. We have also explored Wav2Vec2.0 for multi-stuttered disfluencies, however, Whisper has outperformed in terms of F1-scores and computational efficiency. The micro, macro and weighted F1-scores achieved were 0.88, 0.85, and 0.87 respectively, using Whisper's small variant which only consists of 6 encoder layers. In addition, through our experimental design, we were able to identify the most contributing encoder layer for the task at hand. So by just fine-tuning one encoder layer, we were able to achieve commendable outcomes. Therefore, the learnable parameters of the model were reduced from 20.27 million to just 3.29 million. The innovative approach presented in this paper is an eminent contribution leading to automated assessment of stuttered speech. Looking ahead, we aim to explore its applicability in multilingual cases, along with parameter efficient fine-tuning techniques to make these models more adaptable for different languages and dialects. Nevertheless, a diverse dataset in terms of speakers and disfluency types remains a crucial objective for broader applicability in the real world. 

\section*{Acknowledgement}
\label{sec:acknowlegement}
The authors would like to thank World Technology Partners, USA, for providing support for this study. The CPInS Research Lab at SEECS-NUST and Prince Sultan University have been essential in facilitating the research and publication of this work. The authors also recognize their invaluable support and resources for these efforts.
\section*{Declaration of generative AI and AI-assisted technologies in the writing process}
During the preparation of this paper, the authors used Chat GPT 3.5 in order to rephrase less than 5\% of the paper to improve its readability. After using this service, the authors reviewed and edited the content as needed and take full responsibility for the content of the publication.

\printcredits

%\bibliographystyle{cas-model2-names}
\bibliographystyle{apalike}

\bibliography{cas-refs}


\bio{figs/author1}
HUMA AMEER is a Masters Student at National University of Sciences and Technology (NUST). She received her bachelor’s degree in Computer Sciences from SZABIST, Islamabad. Her research interests include Deep Learning, Speech Processing, and Natural Language Processing.
\endbio

\vspace{1cm} % Add space between the bios

\bio{figs/author2-}
SEEMAB LATIF (Senior Member, IEEE) is an Associate Professor and researcher at National University of Sciences and Technology, NUST, Pakistan. She received her PhD from the University of Manchester, UK. Her research interests includes artificial intelligence, machine learning, data mining and NLP. Her professional services include Industry Consultations, Conference Chair, Technical Program Committee Member and reviewer for several international journals and conferences. In the last 3 years, she has established research collaborations with national and international universities and institutes. She has also secured research grants from National ICT R\&D Grass-Root Initiative, Higher Education Commission Technology Development Fund and UK ILM Ideas. She has received the School Best Teacher award in 2016 and the University Best Innovator Award in 2020. She is also the founder of NUST spin off company, Aawaz AI Tech.
\endbio

\vspace{0.002cm} % Add space between the bios

\bio{figs/author3}
IRAM TARIQ BHATTI has done her Bachelors in Electronics Engineering and Masters in Electrical Engineering from National University of Sciences and Technology (NUST), Islamabad, Pakistan in 2010 and 2019 respectively. She is currently a research associate at CPInS Lab, NUST. Her interest includes applying Deep Learning, Embedded Systems and Formal Verification in EdTech and MedTech. She has won several awards from recognized agencies including International Telecommunication Authority (ITU) in 2012, Pakistan Telecommunication Authority (PTA) in 2015 and attended US State department flagship program International Visitor Leadership Program (IVLP) in 2016.
\endbio

\pagebreak

\vspace{0.002cm} % Add space between the bios

\bio{figs/author4}
RABIA LATIF is currently working as an Assistant Professor at College of Computer and Information Sciences, Prince Sultan University, Riyadh, Saudi Arabia. She received her bachelor’s degree in Computer Science from COMSATS Institute of Information Technology, Islamabad, and master’s degree in information security from National University of Sciences and Technology (NUST), Pakistan. She has done her Ph.D. in cloud-assisted wireless body area networks at National University of Sciences and Technology, Pakistan. Her research interests include wireless body area networks, cloud computing and information security.
\endbio

\end{document}